\documentclass{revtex4}
\usepackage{graphicx}
\usepackage[T1]{fontenc}
\usepackage{amsmath}
\usepackage{amssymb}
\begin{document}

\title{Charge density wave and superconductivity properties in single crystal of Lu$_5$Ir$_4$Si$_{10} $}

\author{Maxime Leroux, Pierre Rodi\`ere, Christine Opagiste }

\address{Institut N\'eel, CNRS et Universit\'e Joseph Fourier, BP 166, F-38042 Grenoble Cedex 9, France}

\begin{abstract}

We measured the electrical resistivity from 2K up to 900K on high quality single crystals of Lu$_5$Ir$_4$Si$_{10} $. A clear thermal hysteresis was found at the onset of the Charge Density Wave (CDW), evidencing the first order nature of the transition. When tantalum is included in the compound, the CDW is destroyed and the superconducting critical temperature is enhanced. Finally, we present specific heat and magnetic penetration depth in the Meissner state. We show that the superconducting properties are very close to a weak coupling BCS superconductor. 

\end{abstract}
\maketitle
\section{Introduction}
\label{intro}
 In many strongly correlated electronic systems, superconductivity seems to be linked to the vicinity of an electronic or magnetic instability with a characteristic phase diagram. Currently, there is a growing interest on the systems where this electronic instability is a charge density wave.  Superconductivity and CDW share the electron-phonon interaction in terms of microscopical mechanism. These kind of competitions were also revived with the recent and quasi-systematic observation of a superconducting pocket at the border of the CDW in the dichalcogenides [1,2]. These superconductors can show deviations from the weak coupling BCS theory, which raises the question of the interplay between the two orders. The CDW opens a gap on parts of the Fermi Surface and can be at the origin of the modification of the superconducting gap structure. This question was mainly asked in the dichalcogenides systems where the CDW is two dimensional (2D). It is important to study other systems where CDW coexist with superconductivity. Here we focus on a compound, Lu$_5$Ir$_4$Si$_{10} $, which offers a completely different type of CDW: 1D, 1st order, and involves a large part of the Fermi Surface. 

Lu$_5$Ir$_4$Si$_{10} $ (P4/mbm tetragonal space group) is constituted of 1D chains of first neighbors lutetium atoms along the c-axis, with a strong interchain coupling.  When the temperature is decreased below T$_{CDW}$, a large jump of the electrical resistivity signs that a large part of the Fermi surface is involved in the CDW.  Moreover, from specific heat measurements, the transition has been suggested to be first order [3]. But, the origin of this first order transition is still unclear, and so far, a full thermal hysteresis at T$_{CDW}$ has not observed. The periodic lattice distortion associated with the CDW manifests itself by the formation of x-ray satellites reflections at wave vectors G+/-n.Q$_{CDW}$ where G is a reciprocal lattice vector, n is an integer and Q$_{CDW}$=(0,0, 3/7). The temperature dependence of the integrated intensity of the satellites reflections peaks, which is related to the CDW order parameter, deviates strongly from a BCS-like dependence [3]. Recently, the origin of the CDW has been attributed to the Peierls mechanism [4].

In this paper, we present electrical resistivity measurements up to 900K on high quality single crystals. We show a full hysteresis at the CDW transition evidencing the first order nature of the transition and the absence of any other CDW transitions at higher temperature. 
We also show that the CDW and superconductivity compete for the same electrons. Finally, we present measurements of specific heat and magnetic penetration depth in the superconducting state.

\section{Experimental method}
\label{sec:1}
A polycrystalline sample of Lu$_5$Ir$_4$Si$_{10} $ was prepared in an induction furnace, with a cold copper crucible and a high-purity argon atmosphere. We used a stoichiometric amount of Lu (99.96\% Ames Lab., or 98\% Alfa Aesar), Ir (99.9\% Alfa Aesar) and Si (99.9999\% Alfa Aesar). Single crystals of Lu$_5$Ir$_4$Si$_{10} $ were grown by two different techniques. On the one hand, a quick thermal quench of the melted polycrystalline sample favored the spontaneous growth of high purity “whiskers” oriented along the c-axis (in the following we will use “W” to denote these samples ) [5]. On the other hand, large single crystals were also grown in a three-arc furnace by a Czochralski technique following the procedure previously published [3]. These latter samples were cut with a wire saw along the crystallographic axes, and subsequently annealed at 900C for 7 days under ultra-high vacuum to reduce the internal stress (we will denote them as “BAR”). 
For the electrical resistivity measurements, we used a traditional four wire technique, with platinum or silver paste to ensure the electrical contacts. The high temperature resistivity was measured in a homemade furnace under vacuum, with a thermometer close to the alumina sample holder. 

The specific heat measurements down to 0.4K were performed using the usual relaxation technique in a commercial PPMS instrument from Quantum Design. Finally, the magnetic penetration depth was measured with a high stability LC oscillator, operating at 14 MHz, driven by a tunnel diode. The superconducting sample is placed at the center of the excitation coil. A small variation of the magnetic penetration depth induces a proportional variation of the inductance and so of the resonant frequency. The calibration is done thanks to the dimension of the sample and by extracting the sample from the excitation coil at low temperature. The experimental set-up is described elsewhere [6]

\section{Charge Density Wave and high temperature resistivity}

\begin{figure}
\includegraphics[width=\textwidth ]{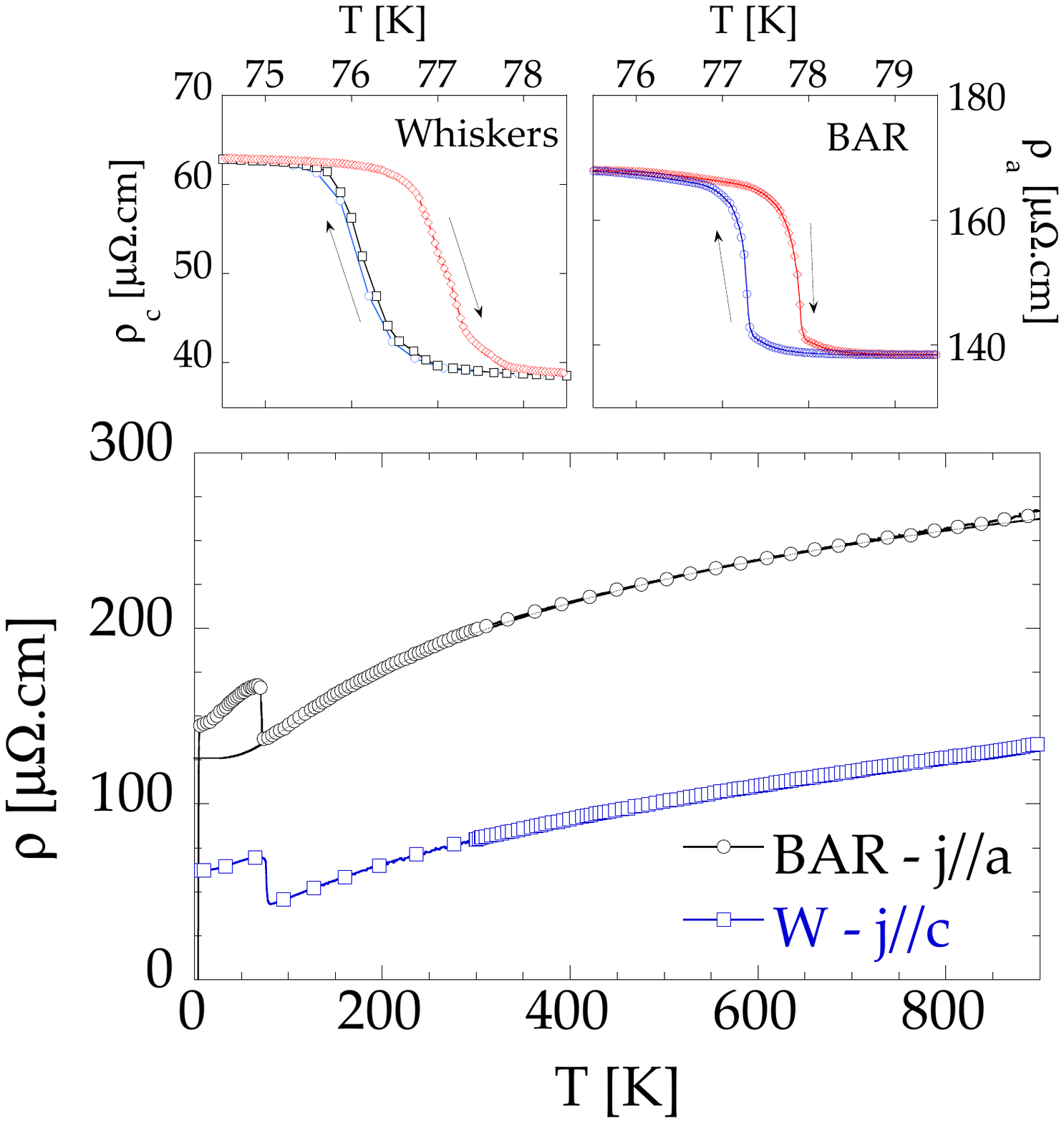}
\caption{High temperature resistivity of a whisker and a bar up to 900K. A fit of the resistivity is shown for j//a (see text). At the top, zoom on the hysteresis behavior at the charge density wave transition for different temperature variation rate (-50mK/min and +/-500mK/min for the whisker, +/-15mK/min for the BAR)}
\label{fig:1}       
\end{figure} 

Fig. 1 shows the resistivity up to 900K for a whisker (where j//c) and for a bar aligned along the a-axis. At low temperature, for the whiskers (respectively BAR) a sharp superconducting transition is measured: TSC =4.05 K (resp. 4.6K) at 50\% of the normal state resistivity. The width of the transition is 50mK (resp. 200mK), with a criterion of 10\% - 90\% of the normal state resistivity. Above T$_{CDW}$~77K, a sudden drop of the resistivity signs the increase of the density of states due to the destruction of the CDW. To check the order of the transition, we increased and decreased the temperature at the vicinity of the CDW at different speed.

A clear thermal hysteresis of the resistivity $\Delta$T~1K, independent of the speed rate, evidences the first order nature of the transition. A thermal hysteresis with $\Delta$T~3K had already been reported in the iso-structural compound Lu$_5$Rh$_4$Si$_{10}$ at the vicinity of the CDW transition [7]. Usually, a first order transition is attributed to a lock-in transition with the lattice of a pre-existing incommensurate CDW at larger temperature, potentially above the room temperature [8].
To check the presence of this possible CDW, we performed resistivity measurements up to 900K. No anomaly on the resistivity is observed. The data are well fitted with a phenomenological model with two parallel resistors [9], one corresponding to the usual resistivity due to impurity and electron-phonon scatterings (Bloch-Grüneisen model), and a second due to saturation when the mean free path l is of the order of the lattice parameter (Ioffe-Regel criterion) : 

\begin{equation}
\rho^{-1}(T)=\rho_{sat}^{-1}+(\rho_{BG}+\rho_0)^{-1}
with \rho_{BG}=4\rho'\Theta^5(\frac{T}{\Theta})\int^{\Theta/T}_0\frac{x^5dx}{(e^x-1)(1-e^{-x}-1)}
\end{equation}
Here $\Theta$ is the Debye temperature and $\rho_{BG}$ is the Bloch Grüneisen formula for an electron-phonon scattering.  An excellent fit is obtained with $\Theta$=315K, and for j//a (resp. j//c) $\rho_{sat}=375\mu\Omega$.cm (resp. 220), $\rho'=0.8\mu\Omega$.cm (resp 0.24) and $\rho_0=200\mu\Omega$.cm (resp. 38). These values are close to the values for other superconducting metals [10].

\section{Competition between Charge density wave and superconductivity} 
A second batch of single crystals was prepared with lutetium including 1\% of tantalum (samples denoted as “Ta”).  Scanning electron microscopy and energy dispersive X-ray spectrometry (EDX) analysis show some small inclusions of pure Ta in the matrix of the Lu$_5$Ir$_4$Si$_{10} $. These inclusions are homogeneously repartitioned in the sample. It is not clear if the Ta has also been substituted to a particular site and/or is in interstitial in the host material. The temperature dependence of the resistivity is compared to pure Lu$_5$Ir$_4$Si$_{10} $ in Fig. 2.

\begin{figure}

  \includegraphics[width=0.8\textwidth, bb= 60 280 500 560]{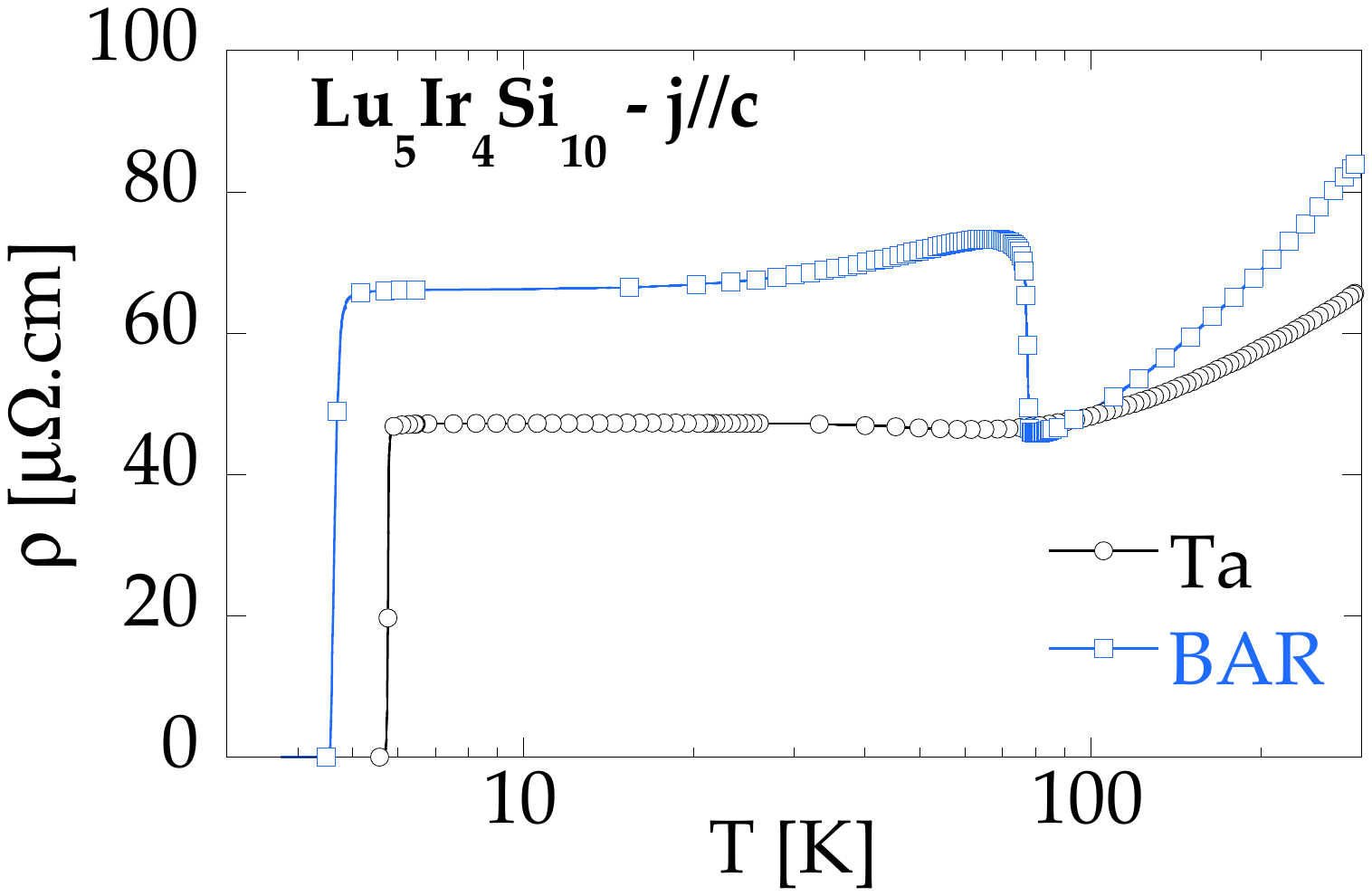}

\caption{Temperature dependence of the resistivity along the c-axis for pure Lu$_5$Ir$_4$Si$_{10} $, compared to the sample with inclusion of tantalum. }
\label{fig:2}       
\end{figure}

The jump of resistivity due to the CDW is totally suppressed. Around 70K, a small minimum of resistivity is reminiscent of the CDW in the pure compound, this suggests the possible presence of a strongly disordered CDW in some parts of the samples. Here, the mechanism at the origin of the suppression of CDW ordering temperature could be a chemical pressure induced by the inclusions of Ta. Further analysis should be performed to clarify this point. The most surprising is the strong increase of the critical temperature of the superconducting transition: from 4.6K in the pure compound to 5.7K. A similar effect has already been clearly reported by substituting Si by Ge [11], or under pressure [12]. This further suggests that both orders are in competition for the same electrons.

\section{Superconducting properties}
To get further insight regarding the superconducting properties of Lu$_5$Ir$_4$Si$_{10} $, we performed specific heat measurement at 0 magnetic field. In the normal state, the specific heat can be fitted with a Sommerfeld coefficient  $\gamma_N$ of 22mJ/mol/K$^2$ and a Debye temperature of 320K, in good agreement with the previous report [12]. This Sommerfeld coefficient corresponds to a density of states at the Fermi level of 0.53 eV$^{-1}$.\AA$^{-3}$ per atom. In Fig. 3 we show the specific heat after subtraction of the normal state contribution. The critical temperature is found to be 3.75K with a width of 120 mK signing a good homogeneity. The temperature dependence of the specific heat in the superconducting state is well described by a weak coupling BCS theory.

\begin{figure}

  \includegraphics[width=0.8\textwidth, bb= 55 360 500 780 ]{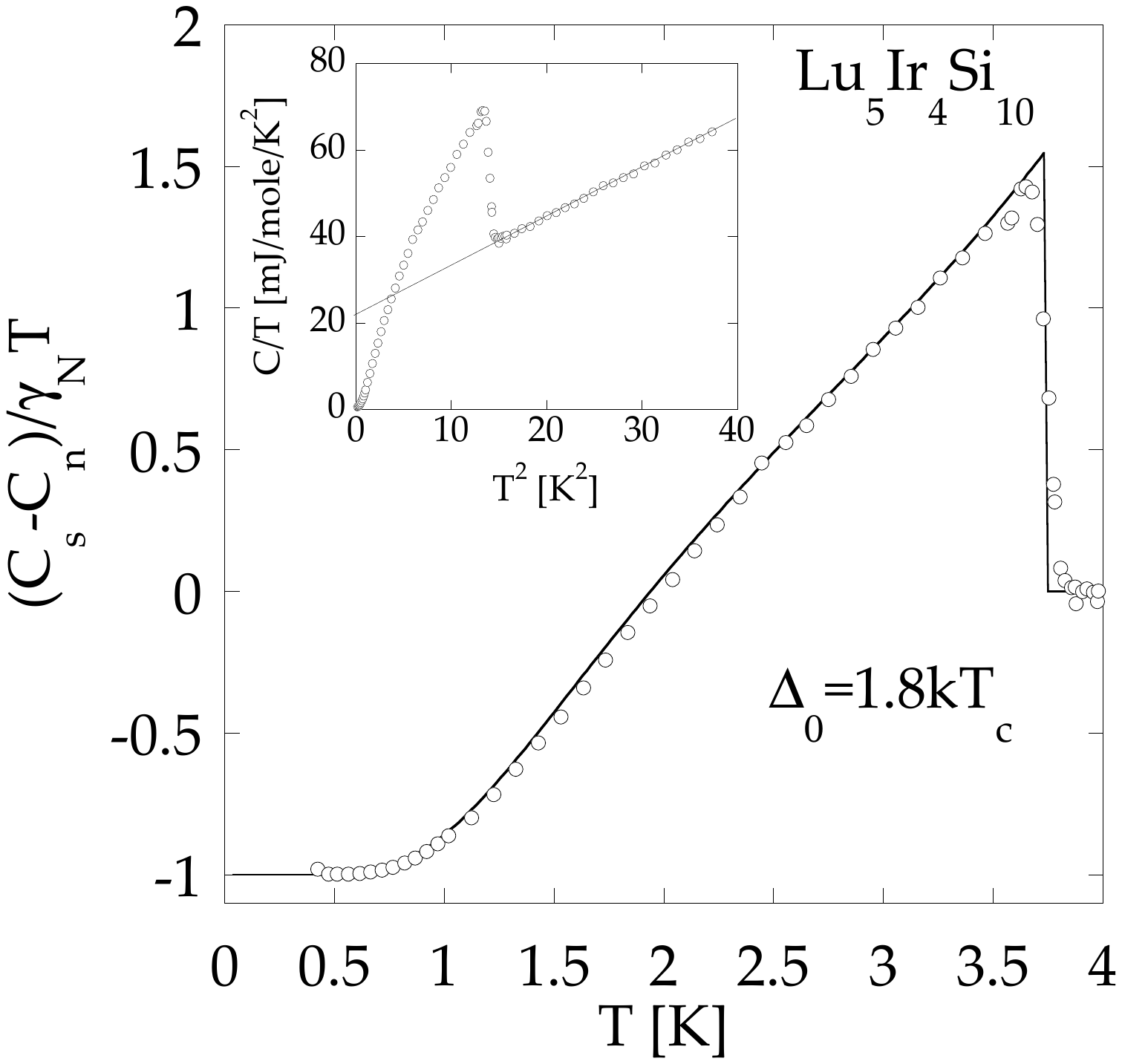}

\caption{Electronic specific heat in the superconducting state. The fit is for a superconducting gap of 1.8kTc. Inset, C/T vs T$^2$ a linear fit in the normal state give the Sommerfeld coefficient and the Debye temperature. }
\label{fig:3}       
\end{figure}

This behavior is strongly different to other system where both orders coexist. For example, 2H-NbSe$_2$ shows the coexistence of a two dimensional (2D) charge density wave (CDW – T$_{CDW}$~33K ) and superconductivity (Tsc~7K). This superconductor shows deviations from the weak coupling BCS theory, which raises the question of the interplay between the two orders [13]. Since these deviations are also present in the iso-structural system 2H-NbS$_2$ despite the absence of CDW [14], it turns out they could come from a peculiar electron-phonon anisotropy general to the transition metal dichalcogenide family.

\begin{figure}

  \includegraphics[width=0.8\textwidth, bb = 30 350 490 790 ]{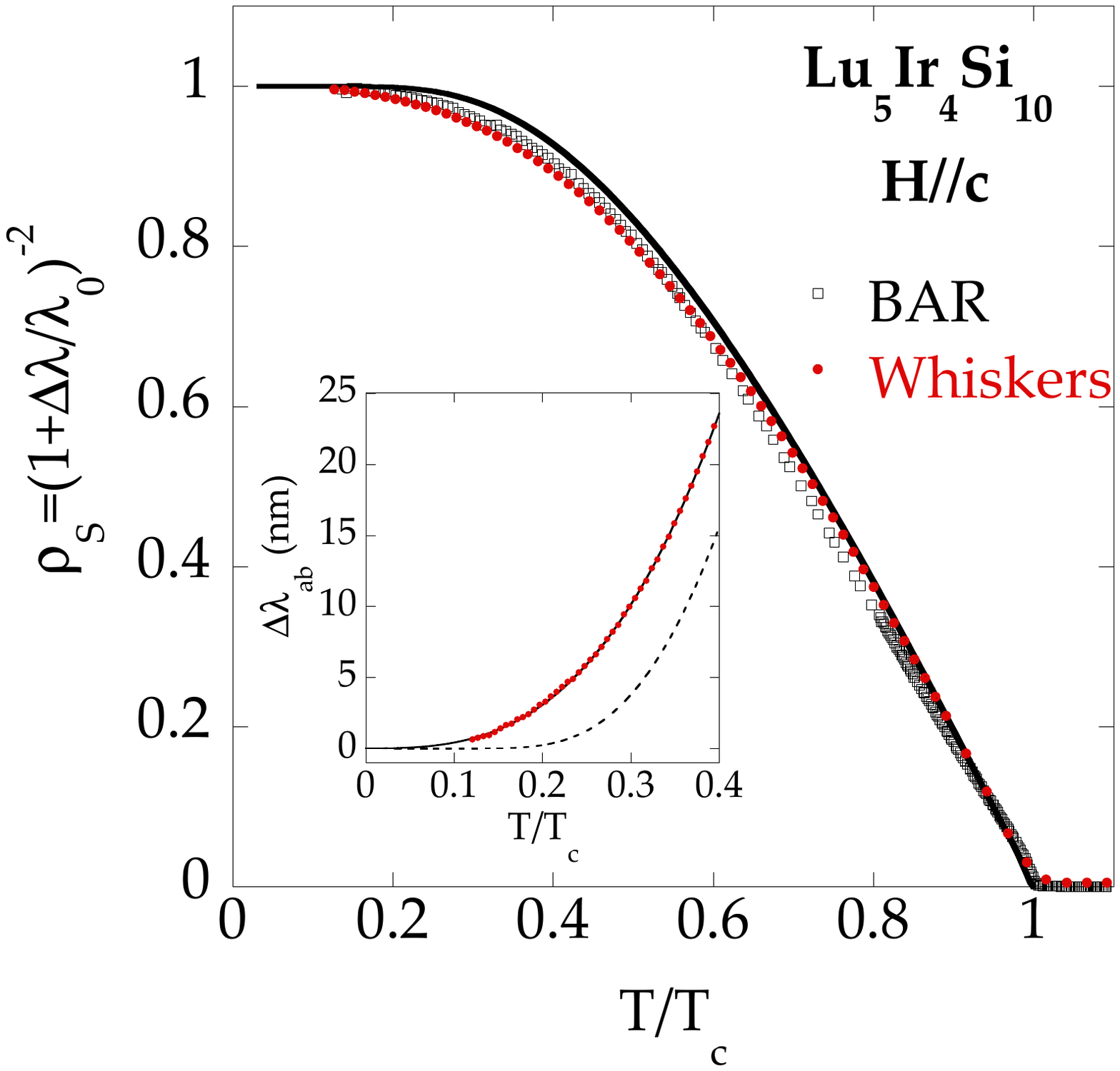}

\caption{Basal normalized superfluid density for a bar (black open square) and a whiskers (closed red circle) with H//c, compared to the weak coupling BCS theory. Here, the critical temperatures are respectively 3.9K for the whisker and 3.73K for the bar. Inset, temperature dependence of the magnetic penetration depth in red, the dashed line is the BCS temperature expected for $\lambda_0$=409nm and $\Delta$=1.76kTc. The solid line is a power law fit.  }
\label{fig:4}       
\end{figure}

Finally, the temperature dependence of the variation of the magnetic penetration depth $\Delta\lambda$(T) was measured in the Meissner state for an AC-magnetic field applied along the c-axis. In this geometry the supercurrents are flowing only in the basal plane. The superfluid density pairs is proportional to $\rho_{S,i}= (\lambda_{0,i}+\lambda_i$(T))$^{-2}$ where i refers to the crystallographic direction, $\lambda_0$ to the absolute value of the magnetic penetration depth at 0K and $\Delta\lambda$ to its variation with temperature. $\lambda_0$ is related only to the Fermi surface. Whereas, the temperature dependence $\Delta\lambda$ is also related to the superconducting gap.  The low temperature approximation for a type II superconductor with a fully open gap is given by : 
\begin{equation}
\Delta\lambda(T)=\lambda_0\sqrt{\frac{\pi}{2}\frac{\Delta_0}{k_B T}}e^{-\frac{\Delta_0}{k_BT}}
\end{equation}
where 0 is the superconducting gap, $k_B$ the Boltzmann constant and $\lambda_0$ the magnetic penetration depth at 0K. In Fig 4. we show a typical temperature dependence of the superfluid density for a single crystal and a whisker, using $\lambda_{0,c}$ of 409nm [15]. All the samples measured show the same temperature dependence.  On a wide temperature range below Tc, the dependence is very close to the one expected for a weak coupling BCS superconductor in the clean limit. This is surprising, because the system is in the dirty limit according to the magnetic penetration depth and the residual resistivity values. Moreover at low temperature we observe a small, but clear deviation to the clean or dirty limit of the  BCS theory. Indeed up to 0.4Tc, the temperature dependence of the magnetic penetration depth is better described by a power law $\beta T^{\alpha}$ with $\alpha$ close to 3 (best fit : $\alpha$=2.94(1) and $\beta$=350(5)nm). The origin of these discrepancies needs further investigations. 

To conclude, we have presented high temperature electrical resistivity showing a clear thermal hysteresis at the vicinity of the CDW transition. It is a strong evidence of the first order nature of the transition. Up to 900K, no anomalies are found. This result suggests that the first order nature of the transition has another origin than in the case of two-dimensionnal dichalcogenides. The superconducting properties are close to a weak coupling BCS theory at the opposite of the dichalcogenides 2H-NbSe$_2$ or 2H-NbS$_2$ which are at the vicinity of a second order CDW transition.\\ 
 
[1] Morosan et al;  Nature Phys. 2 (2006) 544

[2] Sipos et al; Nature Mat. 7 (2008) 960

[3] Becker et al; Phys. Rev. B 59 (1999) 7266

[4] Mansart et al.; PNAS 109 (2012) 5603

[5] Opagiste et al; Journal of Crystal Growth 312 (2010) 3204–3208

[6] Diener et al; Phys. Rev. B 79 220508R (2009)

[7] Lue et al; Phys. Rev. B 66, 033101 (2002)

[8] W. L. McMillan; Phys. Rev. B 12, 1187 (1975)

[9] Wiesmann et al; Phys. Rev. Lett. 38 (1977) 782

[10] Pallechi et al ; Phys. Rev. B 79, 134508  2009

[11] Singh et al; Phys. Rev. B 72 045106 (2005)

[12] Shelton et al; Phys. Rev. B 34 4590 (1986) 

[13] Fletcher et al; PRL 98, 057003 (2007)

[14]  Diener et al; Phys. Rev. B 84, 054531 (2011)

[15] Jaizwal et al; Physica B 312–313 (2002) 142–144




\end{document}